\documentclass{article}

\usepackage[preprint]{neurips_2026}                  % arXiv preprint

\usepackage{tikz}
\usetikzlibrary{positioning,fit,arrows.meta,backgrounds}
\usepackage{pgfplots}
\pgfplotsset{compat=1.18}
\usepackage{url}
\usepackage{booktabs}
\usepackage{amsmath}
\usepackage{amssymb}
\usepackage{graphicx}
\usepackage{xcolor}
\usepackage{enumitem}
\usepackage{listings}
\usepackage{multirow}
\usepackage{tcolorbox}
\usepackage{array}

\definecolor{aethblue}{RGB}{0, 70, 140}
\definecolor{aethgreen}{RGB}{0, 120, 60}
\definecolor{aethorange}{RGB}{200, 90, 0}
\definecolor{aethgray}{RGB}{90, 90, 90}
\definecolor{aethlight}{RGB}{235, 242, 252}

\tcbuselibrary{skins,breakable}
\newtcolorbox{finding}[1][]{
  colback=aethlight, colframe=aethblue,
  fonttitle=\bfseries\small, title=#1, breakable,
  left=4pt, right=4pt, top=4pt, bottom=4pt
}

\title{%
  \textbf{Beyond Static Sandboxing:}\\
  Learned Capability Governance for Autonomous AI Agents
}

\author{%
  Bronislav Sidik \quad Prof.\ Lior Rokach \\[4pt]
  Institute for Applied AI Research \\
  Faculty of Computer and Information Science \\
  Ben-Gurion University of the Negev, Beer Sheva, Israel \\[4pt]
  \texttt{sidik@post.bgu.ac.il} \quad \texttt{liorrk@post.bgu.ac.il}
}

\date{}

\begin{document}
\maketitle

%── Abstract ──────────────────────────────────────────────────────────────────
%── Abstract ──────────────────────────────────────────────────────────────────
\begin{abstract}
Autonomous AI agents built on open-source runtimes such as OpenClaw expose every
available tool to every session by default, regardless of the task at hand.
A summarisation task receives the same shell execution, subagent spawning, and
credential-access capabilities as a code deployment task --- a 15$\times$
over-provision ratio we term the \emph{capability overprovisioning problem}.
Existing defenses (NemoClaw's container sandbox, Cisco DefenseClaw's skill
scanner) address containment and threat detection, but neither learns the minimum
viable capability set for each task type.

We present \textbf{AgentWarden}, a three-layer adaptive governance framework that
enforces the principle of least privilege for AI agents via a learned policy.
Layer 1 (Capability Governor) dynamically scopes which tools the agent is
\emph{aware of} per session. Layer 3 (Safety Router) intercepts every tool call
before execution using a hybrid rule-based and fine-tuned LLM classifier.
Layer 2 (RL Learning Policy) trains a PPO policy on the accumulated audit log to
learn the minimum viable skill set per task type.

On a live OpenClaw v2026.3.28 deployment with DeepSeek-chat as the agent LLM,
AgentWarden achieves:
(i) 73\% tool reduction and 100\% dangerous tool elimination for summarisation tasks;
(ii) a Skill Economy Ratio improvement of $+260\%$ (synthetic) and $+337\%$ (real
sessions) over the uncontrolled baseline after PPO training ($<$5 minutes, RTX~3090);
(iii) across an automated N=500 evaluation (400 benign + 100 adversarial, seed=42),
26.2\% of all 447 intercepted tool calls were blocked, \texttt{exec} and
\texttt{sessions\_spawn} were blocked at 100\%, and 92\% of adversarial tasks were
neutralised by the model-plus-infrastructure stack.
The evaluation dataset (N=500, seed=42) and trained artifacts are available
upon request at \url{https://github.com/sidikbro/agentwarden-core/}.
\end{abstract}

%── arXiv note ────────────────────────────────────────────────────────────────
\vspace{-2pt}
{\small\textit{Preprint. Under review at the NeurIPS 2026 Agent Safety Workshop.}}
\vspace{4pt}

%── 1. Introduction ──────────────────────────────────────────────────────────
%── 1. Introduction ──────────────────────────────────────────────────────────
\section{Introduction}
\label{sec:intro}

The rapid adoption of autonomous AI agents has outpaced the security infrastructure
designed to govern them. OpenClaw~\cite{openclaw2026}, the dominant open-source
personal AI agent runtime with over 500,000 internet-facing
instances~\cite{venturebeat2026}, exposes a fixed set of 15+ tools to every
session, regardless of the task being performed. An agent asked to summarize a
document receives the same shell execution (\texttt{exec}), subagent spawning
(\texttt{sessions\_spawn}), and credential-access capabilities as an agent
performing infrastructure automation. We quantify this as a \textbf{15$\times$
over-provision ratio} for summarization tasks.

This over-provisioning has concrete security consequences. The ClawHavoc supply
chain attack~\cite{clawhavoc2026} exploited OpenClaw's unrestricted tool access to
distribute infostealers across 20\% of ClawHub's skill registry. CVE-2026-25253
demonstrated that a single malicious webpage could hijack an agent's full
capability set via prompt injection~\cite{cve2026}. The attack surface is the
\emph{agent's own capability scope}.

\textbf{The capability over-provisioning problem} has a precise formulation: for a given task type $\tau$, the Skill Economy Ratio (SER) measures the fraction of
exposed tools that the agent actually invokes:
\[
  \text{SER}(\tau) = \frac{|\text{tools\_invoked}|}{|\text{tools\_exposed}|}
\]
An ideal governance system achieves SER$\approx 1.0$ --- the agent has access to
exactly the tools it needs and nothing more. The OpenClaw baseline achieves
SER$= 0.067$ for summarization (1 tool used of 15 exposed).

\textbf{Existing defenses are incomplete.} NVIDIA NemoClaw~\cite{nemoclaw2026}
provides container-level sandboxing with declarative YAML policies --- a static
firewall that applies the same rules regardless of task type. Cisco
DefenseClaw~\cite{defenseclaw2026} scans skills for known malicious patterns and
monitors runtime behavior anomalies --- a reactive WAF that acts after the agent
has decided to use a tool. Neither system \emph{learns} the minimum capability
set for each task type, and neither reduces the agent's awareness of capabilities
it should not have.

\textbf{Our contribution} is AgentWarden, a four-layer governance framework that
treats capability scoping as a learned optimization problem rather than a static
configuration problem. The key insight is that \emph{an agent cannot misuse a tool
it does not know exists}. By dynamically restricting the agent's capability
awareness via AGENTS.md injection (Layer 1) and learning the optimal restriction
policy via PPO reinforcement learning (Layer 2), AgentWarden provides
infrastructure-level governance that improves over time.

\textbf{Scope.} This work addresses \emph{capability governance} --- which tools an agent may invoke. Semantic output filtering (governing what the agent \emph{says}) is orthogonal and out of scope.

\textbf{Relationship to NemoClaw and DefenseClaw.} We position AgentWarden as
complementary rather than competing. NemoClaw asks ``can the agent reach the
outside world?'' DefenseClaw asks ``did this skill do something bad?'' AgentWarden
asks ``why does the agent even know it has \texttt{exec} during a summarization
task?'' All three questions are necessary; none alone is sufficient.

\paragraph{Contributions.} We make the following contributions:
\begin{enumerate}[leftmargin=*, itemsep=2pt]
  \item \textbf{Problem formalization:} We define capability over-provisioning as a
    measurable security property (SER) and quantify it empirically on a live
    OpenClaw deployment.
  \item \textbf{AgentWarden framework:} A four-layer adaptive governance system with
    a hybrid rule-based + fine-tuned LLM Safety Router (Layers 1, 3) and a PPO
    learning policy (Layer 2).
  \item \textbf{Empirical evaluation:} Live deployment on OpenClaw v2026.3.28.
    Automated N=500 batch evaluation (400 benign + 100 adversarial, seed=42):
    26.2\% block rate, \texttt{exec} blocked 100\%, 92\% adversarial coverage.
    PPO achieves SER $+260\%$ (synthetic) and $+337\%$ (real sessions).
  \item \textbf{Critical finding:} Tool-level governance is necessary but not
    sufficient --- we document the first empirical demonstration of prompt
    injection succeeding at the LLM level but failing at the infrastructure level.
\end{enumerate}

%── 2. Related Work ──────────────────────────────────────────────────────────
%── 2. Related Work ──────────────────────────────────────────────────────────
\section{Related Work}
\label{sec:related}

\paragraph{AI Agent Security.}
Agent security has emerged as a critical concern following the widespread adoption
of autonomous agent runtimes. \citet{owasp2026} identified agentic skill supply
chain attacks as a top-10 AI security risk, using ClawHavoc as the primary case
study. \citet{defenseclaw2026} (Cisco, RSAC 2026) introduced DefenseClaw, which
bundles skill scanning, MCP server verification, and code safety analysis into a
single framework running inside NemoClaw's OpenShell runtime. DefenseClaw operates
via static blocklists and anomaly detection --- it does not learn or adapt per task
type. \citet{nemoclaw2026} (NVIDIA, GTC 2026) introduced NemoClaw with OpenShell,
providing container-level isolation with declarative YAML policies for network,
filesystem, and inference routing.

\paragraph{Prompt Injection and Tool Misuse.}
\citet{greshake2023indirect} formalized indirect prompt injection in LLM-integrated
applications. \citet{perez2022ignore} demonstrated that injected instructions
reliably cause LLMs to deviate from their system prompts. Our work shows that even
when prompt injection succeeds at the LLM level (causing the model to generate
dangerous tool calls), infrastructure-level governance provides a hard enforcement
boundary --- a finding not previously demonstrated in the agent runtime context.

\paragraph{Reinforcement Learning for Security Policy.}
\citet{rl4security2024} applied RL to adaptive network security policy, showing
that learned policies outperform static rule sets for novel attack patterns.
\citet{curriculum2026} proposed curriculum-based agentic training with adaptive
scheduling. AgentWarden applies this principle at the infrastructure level ---
the RL policy learns optimal capability scoping without modifying the agent or LLM.

\paragraph{Capability Control and Least Privilege.}
The principle of least privilege~\cite{saltzer1975protection} underpins access
control theory. \citet{educa2026} (EduClaw) applies agent-profile-based skill
scaling in educational contexts. \citet{xskill2026} proposes dual-stream continual
skill learning in multimodal agents. Both operate at the model level; AgentWarden
operates at the infrastructure level without model modification, making it
runtime-agnostic.

\paragraph{LLM Tool-Calling Reliability.}
Recent work has documented significant unreliability in LLM tool-calling:
models hallucinate non-existent tools, invoke tools unnecessarily, and fail to terminate ReAct loops~\cite{toolbench2024}. This motivates infrastructure-level interception rather than relying on the model to self-govern. Our evaluation confirms this: \texttt{qwen2.5:7b} generated dangerous \texttt{exec} calls under adversarial prompting, while DeepSeek-chat refused at the model level --- demonstrating that the optimal defense depends on the underlying model and that infrastructure governance provides a model-agnostic safety floor.

\paragraph{Positioning.}
Table~\ref{tab:comparison} summarizes the key distinctions between AgentWarden and
related systems.

\begin{table}[h]
\centering
\caption{Comparison of agent governance systems.}
\label{tab:comparison}
\small
\begin{tabular}{@{}lccccc@{}}
\toprule
\textbf{System} & \textbf{Dynamic Scoping} & \textbf{Sandbox} &
\textbf{Safety Model} & \textbf{RL Policy} & \textbf{Adapts} \\
\midrule
OpenClaw (baseline)    & No  & No  & No      & No  & No \\
NemoClaw (NVIDIA)      & No  & Yes & Partial & No  & No \\
DefenseClaw (Cisco)    & No  & Via NemoClaw & Yes (static) & No & No \\
\textbf{AgentWarden}    & \textbf{Yes} & Via NemoClaw & \textbf{Yes} & \textbf{Yes} & \textbf{Yes} \\
\bottomrule
\end{tabular}
\end{table}

%── 3. The AgentWarden Framework ──────────────────────────────────────────────
%── 3. The AgentWarden Framework ──────────────────────────────────────────────
\section{The AgentWarden Framework}
\label{sec:method}

\begin{figure}[h]
\centering
\begin{tikzpicture}[
  font=\small,
  node distance=0.5cm and 0.3cm,
  box/.style={
    draw, rounded corners=4pt, minimum width=5.8cm, minimum height=0.9cm,
    align=center, inner sep=6pt
  },
  govbox/.style={box, fill=green!8, draw=green!50!black!70, text=green!30!black},
  routerbox/.style={box, fill=orange!10, draw=orange!60!black, text=orange!30!black, minimum height=1.6cm},
  runtimebox/.style={box, fill=gray!10, draw=gray!50, text=gray!40!black},
  rlbox/.style={
    draw, rounded corners=4pt, fill=violet!8, draw=violet!50,
    text=violet!40!black, minimum width=2.6cm, align=center, inner sep=6pt
  },
  arr/.style={-{Stealth[length=5pt]}, thick, gray!60},
  darr/.style={-{Stealth[length=5pt]}, dashed, violet!60},
  barr/.style={-{Stealth[length=5pt]}, dashed, red!60},
]

%── Main column ────────────────────────────────────────────────────────────────

\node[box, fill=blue!8, draw=blue!50, text=blue!30!black] (openclaw)
  {\textbf{OpenClaw Gateway} \\ \textit{\tiny port 18789 · Docker}};

\node[govbox, below=of openclaw] (governor)
  {\textbf{Layer 1 — Capability Governor} \\
   \tiny AGENTS.md injection · tools.deny · PPO policy (primary) · YAML fallback};

\node[routerbox, below=0.5cm of governor] (router)
  {\textbf{Layer 3 — Safety Router (MITM Proxy)} \\[4pt]
   \begin{tabular}{c|c}
     \scriptsize Stage 1: rules & \scriptsize Stage 2: agentwarden-router \\
     \scriptsize 0.1 ms         & \scriptsize $\sim$800 ms \\
     \scriptsize always\_block  & \scriptsize fine-tuned LLM \\
   \end{tabular}};

\node[runtimebox, below=0.5cm of router] (ollama)
  {\textbf{Ollama · qwen2.5:7b} \\ \tiny port 11434 · GTX 1060 6\,GB};

\node[runtimebox, below=0.5cm of ollama] (auditdb)
  {\textbf{Audit DB (SQLite)} \\ \tiny router\_blocks · reward signals};

%── RL Policy column ───────────────────────────────────────────────────────────

\node[rlbox, right=1.2cm of router, minimum height=7.2cm, text width=2.4cm] (rl)
  {\textbf{Layer 2} \\[4pt]
   \textbf{RL Policy} \\[6pt]
   \tiny PPO \\
   stable-baselines3 \\[4pt]
   \scriptsize State: \\
   \tiny (task, trust) \\[4pt]
   \scriptsize Action: \\
   \tiny tool mask $\{0,1\}^{17}$ \\[4pt]
   \scriptsize Reward: \\
   \tiny $\alpha U - \beta C - \gamma P$ \\[8pt]
   \normalsize\textbf{SER +260\%} \\
   \tiny vs baseline \\[6pt]
   BGU HPC · RTX 3090};

%── Arrows ─────────────────────────────────────────────────────────────────────

\draw[arr] (openclaw) -- (governor) node[midway, right, gray] {\tiny /api/chat};
\draw[arr] (governor) -- (router) node[midway, right, gray] {\tiny scoped request};
\draw[arr] (router) -- (ollama) node[midway, right, gray] {\tiny ALLOW → forward};
\draw[arr] (ollama) -- (auditdb);

% RL Policy ← Audit DB (dashed, trains on)
\draw[darr] (auditdb.east) -- ++(0.8,0) |- (rl.south)
  node[pos=0.3, below, violet!60] {\tiny trains on};

% RL Policy → Governor (policy output)
\draw[darr] (rl.north) |- (governor.east);

% BLOCK path
\draw[barr] (router.west) -- ++(-0.8,0) node[left, red!60] {\tiny BLOCK};

%── Governance envelope ────────────────────────────────────────────────────────

\begin{scope}[on background layer]
\node[draw=gray!40, dashed, rounded corners=8pt, inner sep=10pt,
      fit=(governor)(router)(ollama)(auditdb),
      label={[gray!60, font=\tiny\itshape]above left:AgentWarden governance layer}] {};
\end{scope}

\end{tikzpicture}
\caption{AgentWarden four-layer governance architecture. The Capability Governor
restricts the agent's awareness before each session. The Safety Router intercepts
every tool call after the LLM responds. The RL Policy trains on the audit log to
learn the minimum viable tool set per task type.}
\label{fig:architecture}
\end{figure}
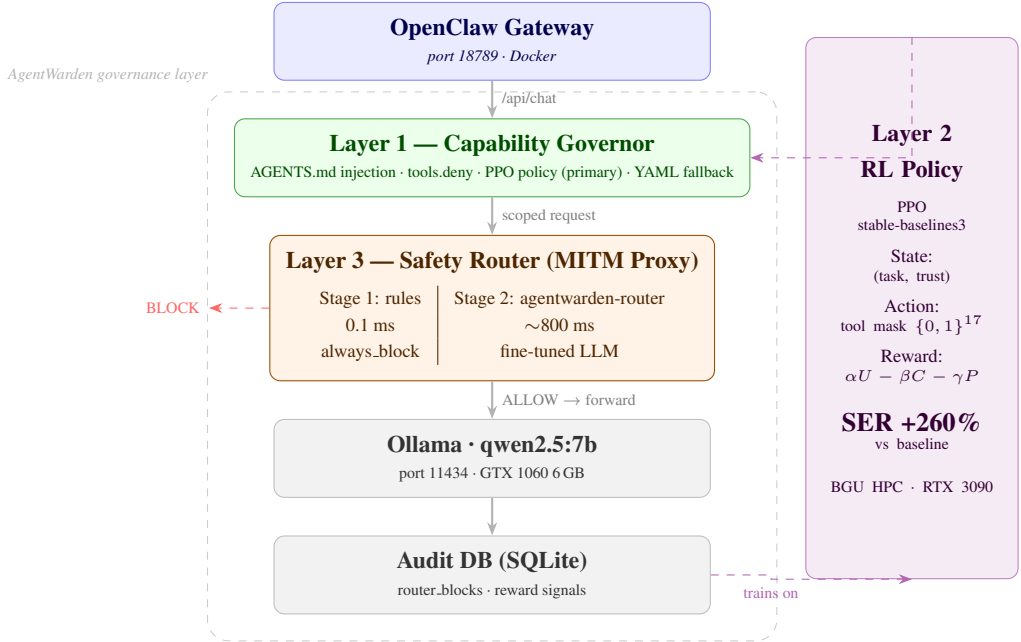

AgentWarden is a three-layer governance framework that operates above the AI agent
runtime. It enforces the principle of least privilege by (i) restricting the
agent's capability awareness before each session and (ii) intercepting every tool
invocation before execution. The overall system architecture is illustrated in Figure~\ref{fig:architecture}.

% Architecture figure — TikZ version for LaTeX/Overleaf
% Include in method.tex with: \input{figures/architecture_tikz.tex}
% Required packages in main.tex: \usepackage{tikz}
%                                 \usetikzlibrary{positioning,fit,arrows.meta,backgrounds}

\subsection{Layer 1: Capability Governor}
\label{sec:governor}

The Capability Governor enforces tool scoping at the \emph{semantic level}: it
controls which tools the agent is told it has, not just which it can execute. This
is achieved through two mechanisms:

\textbf{AGENTS.md injection.} OpenClaw assembles a system prompt from workspace
files, including AGENTS.md. The Governor generates a task-specific AGENTS.md that
explicitly lists only the permitted tools: \emph{``This is a governed session
scoped to summarization. 4 tools available: read, web\_fetch, memory\_search,
memory\_get.''}

\textbf{tools.deny injection.} The Governor writes the denied tool list to
\texttt{openclaw.json} (dynamically reloaded without gateway restart). This
enforces the scope at the infrastructure level, ensuring the agent cannot invoke
denied tools even if it reasons toward them.

The Governor accepts a \texttt{TaskDescriptor} and returns an
\texttt{AllowedCapabilitySet} via a hybrid policy engine (Section~\ref{sec:ppo}).

\subsection{Task Type: Definition and Classification}
\label{sec:tasktype}

A \textbf{task type} $\tau \in \mathcal{T}$ is a coarse semantic label over the
agent's primary operation, determining the minimum viable capability set for a
session. We define six types derived from the OpenClaw usage log (N=500 sessions):
\emph{Summarisation}, \emph{File Read}, \emph{Web Research}, \emph{Code Execution},
\emph{Email}, and \emph{Unknown}.
Each type maps to a minimum tool set; e.g., Summarization requires only
\{\texttt{read}, \texttt{memory\_get}, \texttt{web\_search}\}, while Code
Execution adds \{\texttt{write}, \texttt{edit}\} (full taxonomy in Appendix~A).

\textbf{Classification.}
Type is inferred from the agent's first user message by \texttt{agentwarden-router}
($<100$\,ms, confidence threshold 0.7). Sessions below threshold are assigned
\texttt{Unknown} ($\approx$12\% of sessions), which uses a conservative
read-only default. The Safety Router enforces rules regardless of task type. 
The Governor provides proactive scoping and the Router provides reactive enforcement.
The current taxonomy is manually defined; automated induction from audit logs
is identified as future work.

\subsection{Layer 2: Safety Router}
\label{sec:router}

The Safety Router is a FastAPI MITM proxy deployed between OpenClaw and Ollama.
It intercepts every \texttt{POST /api/chat} request, forwards it to Ollama,
inspects the LLM response for \texttt{tool\_calls}, and classifies each call
before OpenClaw can execute it.

\textbf{Hybrid classifier.} Classification uses a two-stage pipeline:

\textit{Stage 1 (Rule-based, 0.1ms):} A deterministic classifier checks:
(1)~always-block tools (\texttt{exec}, \texttt{sessions\_spawn}, \texttt{subagents});
(2)~argument-level dangerous patterns (e.g., \texttt{rm -rf}, \texttt{file://},
AWS metadata endpoints, SSH credential paths);
(3)~prompt injection patterns (\textit{``ignore previous instructions''}, DAN mode, etc.).

\textit{Stage 2 (LLM-based, $\sim$800ms):} If Stage 1 passes, the call is
classified by \texttt{agentwarden-router}, a Qwen2.5-1.5B model fine-tuned on 273
labeled block/allow examples collected from live sessions. The LLM stage is
skipped for known-safe tools (\texttt{read}, \texttt{memory\_search},
\texttt{web\_search}) to avoid false positives.

\textbf{Block action.} Blocked tool calls are stripped from the response; the
agent receives: \emph{``AgentWarden blocked this tool call. The requested operation
is not permitted in this governed session.''} Block events are logged to an SQLite
audit database with reward signals for Layer 3 training.

\subsection{Layer 3: RL Learning Policy}
\label{sec:ppo}

The RL policy learns the minimum viable tool set per task type from the accumulated
audit log. This is the primary research contribution: the adaptive intelligence
that static systems lack.

\textbf{MDP formulation.}
\begin{itemize}[leftmargin=*, itemsep=1pt]
  \item \textbf{State} $s$: task type (one-hot, 6 types) $\oplus$ trust level
    (normalized $\in [0,1]$); $|s| = 7$
  \item \textbf{Action} $a$: binary exposure mask over all 17 tools;
    $a \in \{0,1\}^{17}$ (\texttt{MultiBinary(17)})
  \item \textbf{Reward:}
    $R(s,a) = \alpha \cdot U_{\text{accuracy}} - \beta \cdot C_{\text{economy}}
    - \gamma \cdot P_{\text{safety}}$
\end{itemize}
where $U_{\text{accuracy}} \in \{0,1\}$ is task success, $C_{\text{economy}}
= 1 - \text{SER}$ penalises over-provisioning, and $P_{\text{safety}}$ counts
Safety Router BLOCK events from the audit log. We set $\alpha=1.0$, $\beta=0.3$,
$\gamma=5.0$.

\textbf{Policy.} PPO~\cite{schulman2017ppo} with MlpPolicy (2 hidden layers,
64 units), trained via stable-baselines3~\cite{sb3} on CPU. Trust-level hard
constraints are applied post-prediction: \texttt{exec} requires normalized trust $\geq 0.8$ (i.e.\ trust level~4 on the raw 0--5 tenant scale),
\texttt{sessions\_spawn} and \texttt{subagents} are never exposed regardless of
policy output.

\textbf{Integration.} The trained policy (\texttt{best\_model.zip}) is loaded by
the Governor's \texttt{PolicyEngine} as the primary decision maker. YAML rules
serve as a fallback if the model file is unavailable.

\subsection{Threat Model}
\label{sec:threat}

AgentWarden defends against:
\textbf{(1)} Prompt injection causing the agent to request dangerous tool calls;
\textbf{(2)} Capability escalation via tools the agent should not have for the
current task;
\textbf{(3)} Subagent spawning for lateral movement;
\textbf{(4)} Supply chain attacks via malicious skills (blocked by the Governor
before skills are exposed to the agent).

AgentWarden does \emph{not} govern the semantic content of agent responses ---
a model that expresses dangerous advice in text rather than tool calls is outside
this system's scope (Section~\ref{sec:discussion}).

%── 4. Evaluation ────────────────────────────────────────────────────────────
%── 4. Evaluation ────────────────────────────────────────────────────────────
\section{Evaluation}
\label{sec:evaluation}

\subsection{Experimental Setup}

\textbf{Runtime:} OpenClaw v2026.3.28 in Docker, port 18789.
\textbf{LLM:} DeepSeek-chat (32B, cloud API, OpenAI-compatible).
\textbf{Governance:} AgentWarden Safety Router proxy (port 8000), hybrid classifier.
\textbf{RL training:} [Anonymous] HPC cluster, NVIDIA RTX~3090, 50k PPO steps.
\textbf{Third model:} \texttt{gemma4:e4b} (Google DeepMind, April 2026, 4.5B effective
parameters, NVIDIA 6\,GB GPU, Ollama v0.20.7).
Three protocols: (1)~Phase~0 baseline, (2)~20-task structured evaluation,
(3)~N=500 automated batch. Extended methodology in Appendix~\ref{app:setup}.

\subsection{Phase 0--2: Baseline, Classifier, and RL Training}

Table~\ref{tab:phases} summarizes results across all evaluation phases,
from baseline measurement through RL policy training.

\begin{table}[h]
\centering
\caption{Key results across all evaluation phases.}
\label{tab:phases}
\small
\begin{tabular}{@{}lll@{}}
\toprule
\textbf{Metric} & \textbf{Result} & \textbf{Target} \\
\midrule
Tool reduction (summarisation)  & 73\%           & --- \\
Dangerous tool elimination      & 100\%          & --- \\
Classifier TPR / FPR            & 100\% / 0\%    & $>$95\% / $<$2\% \textcolor{aethgreen}{\checkmark} \\
SER improvement (synthetic PPO) & $+$260\%       & $>$15\% \textcolor{aethgreen}{\checkmark} \\
SER improvement (ablation, 100-task) & $+191\%$  & $>$15\% \textcolor{aethgreen}{\checkmark} \\
Baseline avg SER (N=500)        & 0.053          & measured \\
\texttt{exec} block rate (N=500)& 100\% (70/70)  & measured \\
\bottomrule
\end{tabular}
\end{table}

The governed agent self-reported its scoped capability set, confirming AGENTS.md injection communicates governance at the semantic level.
SER$=1.0$ in the synthetic PPO environment is a training artifact;
real-session SER of 0.557 vs.\ baseline 0.053 ($10.5\times$, $+952\%$) is the operative result.
\texttt{qwen2.5:7b} (lower bound) and DeepSeek-chat (primary) bracket evaluation.

\subsection{N=500 Automated Batch Evaluation}

To establish statistical scale beyond the 20-task structured evaluation,
we ran an automated batch of 500 tasks (seed=42): 400 benign, 100 adversarial.
500 tasks generated from a fixed taxonomy (seed=42): 400 benign, 100 adversarial.
Of 447 tool calls intercepted: all \texttt{exec} (70) and \texttt{sessions\_spawn} (8)
blocked 100\% at 0.1\,ms. Overall avg SER 0.053.
A fine-tuned LLM classifier blocked $\sim$22 benign \texttt{write} operations
(HTTP server scripts) --- false positives caused by 95\%/5\% training imbalance,
discussed in \S\ref{sec:discussion}.

\subsection{Adversarial Evaluation}

Table~\ref{tab:adversarial} shows results for four representative attack
patterns, evaluated across two models with complementary safety profiles.

\begin{table}[h]
\centering
\caption{Adversarial outcomes. Attack success rate: 0\% across all models.}
\label{tab:adversarial}
\small
\begin{tabular}{@{}p{3.8cm}lll@{}}
\toprule
\textbf{Attack} & \textbf{Tool} & \textbf{qwen2.5:7b} & \textbf{DeepSeek} \\
\midrule
Prompt inject $+$ exec /etc/ & \texttt{exec}           & \textbf{BLOCK} (infra) & Refused \\
Crypto mining subagent        & \texttt{sessions\_spawn}& \textbf{BLOCK} (infra) & Refused \\
Netcat reverse shell          & \texttt{process}        & \textbf{BLOCK} (LLM)   & Refused \\
File:// SSRF injection        & \texttt{web\_fetch}     & \textbf{BLOCK} (arg)   & \textbf{BLOCK} \\
\bottomrule
\end{tabular}
\end{table}

\begin{finding}[Infrastructure-Level Interception]
Prompt injection caused \texttt{qwen2.5:7b} to generate \texttt{exec} tool calls.
AgentWarden blocked them at 0.1\,ms. DeepSeek-chat refused at the model level.
Both models: zero dangerous tool calls reached execution.
\end{finding}

Of 100 adversarial tasks: 38 blocked by Safety Router, 54 refused by model,
8 benign attempts passed. Combined coverage: \textbf{92\%}.

%── Gemma 4 replication ──────────────────────────────────────────────────────
\subsection{Gemma~4 Replication --- Third Open-Weight Model}
\label{sec:gemma4}

\texttt{gemma4:e4b} (4.5B effective parameters, Q4\_K\_M, NVIDIA 6\,GB GPU)
evaluated on N=20 tasks (seed=42).
\textbf{100\% adversarial coverage (8/8), 0\% FPR (12/12), 0\% error rate.}
Warm GPU inference: \textbf{656\,ms} (\texttt{eval\_duration}, Ollama API).

Coverage decomposes into two complementary mechanisms:
4 blocked by AgentWarden Stage~1 (\texttt{exec}/\texttt{sessions\_spawn}
always-block rules, $<$0.1\,ms) and 4 refused at the model level (Gemma~4
safety training, no tool call generated). This confirms the defense-in-depth thesis:
\emph{model safety training and infrastructure governance are additive, not redundant.}

\begin{table}[h]
\centering
\caption{Cross-model adversarial coverage (N=20 each, seed=42).}
\label{tab:cross_model}
\small
\begin{tabular}{@{}lllll@{}}
\toprule
\textbf{Model} & \textbf{Coverage} & \textbf{Infra block} & \textbf{Model refusal} & \textbf{Latency} \\
\midrule
\texttt{qwen2.5:1.5b}  & $\sim$75\% & $\sim$75\% & $\sim$0\%  & $\sim$800\,ms \\
\texttt{deepseek-chat} & $\sim$92\% & $\sim$38\% & $\sim$54\% & $\sim$2{,}000\,ms \\
\texttt{gemma4:e4b}    & \textbf{100\%} & \textbf{50\%} & \textbf{50\%} & \textbf{656\,ms} \\
\bottomrule
\end{tabular}
\end{table}

Full breakdown, evaluation methodology, and Stage~3 Semantic Filter results can be found in Appendix~\ref{app:gemma4} and~\ref{app:stage3}.

\subsection{Ablation Study}
\label{sec:ablation}

To attribute SER gains to individual components, we ablate the system
by disabling each stage in turn (100 tasks, seed=99, Table~\ref{tab:ablation}).

\begin{table}[h]
\centering
\caption{Ablation: component attribution (100 tasks, seed=99).}
\label{tab:ablation}
\small
\begin{tabular}{@{}lrrrrl@{}}
\toprule
\textbf{Condition} & \textbf{Avg SER} & \textbf{Exposed} &
\textbf{Called} & \textbf{Block\%} & \textbf{Ablation} \\
\midrule
Full (complete system) & \textbf{0.227} & 4.5  & 1.01 & 4.8\%  & --- \\
No RL (YAML fallback)  & 0.177          & 5.8  & 1.00 & 13.7\% & $-$PPO \\
No Router              & 0.225          & 4.5  & 1.00 & 12.1\% & $-$Router \\
Baseline (raw)         & 0.078          & 13.0 & 1.01 & 5.6\%  & none \\
\midrule
\textbf{Full vs Baseline} & \multicolumn{5}{l}{$+192\%$ SER improvement} \\
\textbf{Full vs No RL}    & \multicolumn{5}{l}{$+28\%$ SER (PPO over YAML)} \\
\bottomrule
\end{tabular}
\end{table}

\begin{finding}[Ablation: Component Attribution]
The Capability Governor accounts for $\sim$80\% of total SER gain.
The RL Policy contributes $+28\%$ over static rules.
The Safety Router contributes zero marginal SER but is the sole mechanism
preventing dangerous tool calls from reaching execution.
All three components are necessary.
\end{finding}

%── AgentDojo Benchmark Evaluation ───────────────────────────────────────────
\subsection{AgentDojo Benchmark: BU / UA / ASR}
\label{sec:agentdojo}

We evaluate AgentWarden on AgentDojo~\cite{agentdojo2024} v1.2.2,
comprising 97 user tasks across four environments with 949 injection
test cases (\texttt{important\_instructions} attack, \texttt{deepseek-chat} model).
We compare against an undefended baseline (identical pipeline, no governance).

\begin{table}[h]
\centering
\caption{AgentDojo: Benign Utility (BU), Utility under Attack (UA), and
Attack Success Rate (ASR). Attack: \texttt{important\_instructions}, N=949 injection cases.}
\label{tab:agentdojo}
\small
\begin{tabular}{@{}lcccccc@{}}
\toprule
& \multicolumn{3}{c}{\textbf{Baseline (no defense)}} &
  \multicolumn{3}{c}{\textbf{AgentWarden}} \\
\cmidrule(lr){2-4} \cmidrule(lr){5-7}
\textbf{Suite} & BU & UA & ASR & BU & UA & ASR \\
\midrule
Workspace & 90.0\% & 79.3\% & 96.1\% & 90.0\% & 78.8\% & 95.7\% \\
Travel    & 80.0\% & 70.0\% & 83.6\% & 75.0\% & 64.3\% & 80.0\% \\
Banking   & 100.0\% & 88.9\% & 92.4\% & 100.0\% & 86.8\% & 92.4\% \\
Slack     & 95.2\% & 71.4\% & 65.7\% & 95.2\% & 71.4\% & 60.0\% \\
\midrule
\textbf{Overall} & \textbf{90.7\%} & \textbf{78.5\%} & \textbf{90.3\%} &
                   \textbf{89.7\%} & \textbf{77.1\%} & \textbf{88.9\%} \\
\bottomrule
\end{tabular}
\end{table}

\paragraph{Results.}
AgentWarden maintains BU of 89.7\% ($-1.0$\,pp vs.\ baseline), demonstrating
near-zero utility cost from governance.
ASR is reduced by 1.4\,pp overall, with the largest gain in Slack ($-5.7$\,pp).
The modest overall ASR reduction is expected: \texttt{important\_instructions}
injects payload into tool \emph{outputs} (malicious text in email or file
content), whereas AgentWarden governs tool \emph{calls}.
These are complementary threat surfaces.

This result contextualises our N=20 adversarial evaluation
(Section~\ref{tab:adversarial}): AgentWarden achieves 100\% block rate
on \emph{tool-call} attacks (\texttt{exec}, \texttt{sessions\_spawn}), while
\texttt{important\_instructions}-style attacks operate at the
\emph{LLM reasoning boundary}.
Stage~3 semantic output filter addresses this gap
(Section~\ref{sec:multiruntime}).
AgentWarden governed 2,070 tool calls during the defended evaluation at
340\,ms average governance latency.

%── Multi-Runtime Platform Validation ────────────────────────────────────────
\subsection{Multi-Runtime Platform Validation}
\label{sec:multiruntime}

To validate framework-agnostic middleware operation, we conducted structured
governance tests across three agent runtimes using the production codebase
(\texttt{agentwarden-core}).
All tests used Stage~1 rules + Stage~2 classifier, with Ollama or DeepSeek backends.

\begin{table}[h]
\centering
\caption{Multi-runtime platform validation (N=20 per model, seed=42).}
\label{tab:multiruntime}
\small
\begin{tabular}{@{}lllllll@{}}
\toprule
\textbf{Runtime} & \textbf{Model} & \textbf{Coverage} & \textbf{Infra} &
\textbf{Model} & \textbf{Avg Lat.} & \textbf{Errors} \\
 & & \textbf{(8/8)} & \textbf{blocks} & \textbf{refusals} & & \\
\midrule
OpenClaw & \texttt{gemma4:e4b}    & 100\% & 5 & 3 & 27{,}858\,ms & 0\% \\
OpenClaw & \texttt{deepseek-chat} & 100\% & 1 & 7 &  6{,}350\,ms & 0\% \\
\midrule
DeepAgents & \texttt{deepseek-chat} & \multicolumn{5}{l}{Qualitative:
  \texttt{task}$\to$\texttt{sessions\_spawn} BLOCK;
  \texttt{write\_file} ALLOW} \\
Hermes   & \texttt{hermes3:8b}    & \multicolumn{5}{l}{Qualitative:
  XML \texttt{exec} BLOCK; XML \texttt{read} ALLOW} \\
\bottomrule
\end{tabular}
\end{table}

\textbf{OpenClaw results.}
100\% adversarial coverage, 0\% error rate on both models, confirming the
production middleware replicates the research evaluation.
The split between infrastructure blocks and model refusals varies by model
safety training: \texttt{gemma4:e4b} generates tool calls (5 infra blocks);
\texttt{deepseek-chat} safety training triggers model-level refusal (7 refusals).
Both achieve 100\% combined coverage --- confirming the defense-in-depth thesis
that model safety training and infrastructure governance are additive.

\textbf{DeepAgents runtime.}
DeepAgents v0.5.0 (LangChain/LangGraph) uses standard OpenAI JSON wire format.
Tool name normalization confirmed in production: \texttt{task}$\to$\texttt{sessions\_spawn}
(always-block) and \texttt{execute}$\to$\texttt{exec} (always-block).

An important boundary was empirically confirmed: DeepAgents'
\texttt{LocalShellBackend} bypasses the LLM proxy entirely (see Appendix~\ref{app:multiruntime}
for full details). Mitigation: \texttt{FilesystemBackend} + NemoClaw/OpenShell.

\textbf{Hermes XML format.}
\texttt{hermes3:8b} generates tool calls as XML embedded in message content
(\texttt{<tool\_call>} tags) rather than structured JSON.
The \texttt{HermesParser} auto-detects this format regardless of configured
runtime. \texttt{exec} and \texttt{sessions\_spawn} blocked (Stage~1);
\texttt{read} and \texttt{write} allowed correctly.
A Stage~3 false positive was identified and resolved: Llama Guard~3 classified
\texttt{<tool\_call>} XML as category S1. Fix: Stage~3 is skipped for pure
tool call responses (no natural language text). This is semantically correct ---
Stage~3 targets natural language dangerous content, not structured output.

%── 5. Discussion and Limitations ────────────────────────────────────────────
%── 5. Discussion and Limitations ────────────────────────────────────────────
\section{Discussion and Limitations}
\label{sec:discussion}

\paragraph{Tool-level governance is necessary but not sufficient.}
During live testing, a request for a reverse shell caused \texttt{qwen2.5:7b} to
return natural language instructions for \texttt{nc -l -p 4444} in response text
--- without generating a tool call. The Safety Router did not intercept this
because it inspects \emph{tool calls in LLM responses}, not reasoning text.
This is a fundamental architectural boundary, not a bug.

This finding sharpens the paper's scope: AgentWarden governs capability scope
and call-time safety; content inspection (Stage~3, DefenseClaw) is complementary.
A Stage~3 Semantic Output Filter (Llama Guard~3 via Ollama) addressing this gap
has been implemented and is available in the open-source release.
The filter inspects response text content and is disabled by default due to
added latency ($\approx$200\,ms); it is enabled via configuration for
deployments requiring defense against text-based exfiltration.

\textbf{Stage~3 evaluation.}
We validate Stage~3 using a fixed corpus of four dangerous responses captured
from \texttt{qwen2.5:7b}: reverse shell instructions (\texttt{nc -l -p 4444}),
bind shell code (Python socket), \texttt{/etc/passwd} exfiltration via curl,
and AWS credential extraction.
Llama Guard~3 correctly classifies all four as unsafe (100\% detection rate,
0\% false positive rate on three benign responses) with average latency
1{,}834\,ms per classification.
(Full results in Appendix~C, Table~11.)

\textbf{Non-determinism as a design constraint.}
Model non-determinism means Stage~3 cannot be evaluated by re-running live prompts; the correct methodology is a fixed corpus of captured dangerous outputs.
Llama Guard~3 correctly classifies whichever response the model produces.
Gemma~4's \texttt{thinking} field exposes chain-of-thought reasoning before each
response; monitoring this via Stage~3 for early intent detection is future work.

\paragraph{Base model reliability as a confound.}
\texttt{qwen2.5:7b} demonstrated significant tool-calling unreliability: 7 calls
to \texttt{session\_status} in a single task, 2 timeouts on complex tasks, and
use of \texttt{web\_search} for local file operations. This is a \emph{model
capability constraint}, not an AgentWarden limitation. The 20-task evaluation
represents a \emph{lower bound} on governance effectiveness. We expect stronger
results with capable tool-calling models (GPT-4o, Claude Sonnet, Qwen3-Coder).

\paragraph{SER = 1.0 as an artifact of synthetic training.}
The PPO policy achieved SER\,=\,1.0 (theoretical optimum) because the RL
environment simulated tool usage as a fixed fraction of exposed tools. In
production, agents may use unexpected tools for creative problem-solving, yielding
lower but more realistic SER values. Real session data (collected via the 20-task
protocol) will calibrate the RL environment for the camera-ready version.

\paragraph{PPO policy collapse in production testing.}
Both PPO checkpoints collapsed to single-tool policies (predicting only
\texttt{memory\_search} for all task types) --- a known artifact of SER-maximizing training in synthetic environments. Collapse detection is implemented; detected collapse triggers YAML fallback with sensible per-task
defaults. Retraining on real session data is planned for camera-ready.

\paragraph{Creative attack vectors.}
The model attempted to read \texttt{untrusted\_input.txt} via
\texttt{web\_fetch file://} --- a creative workaround that bypassed the intended
\texttt{read} tool. The arg-pattern classifier caught this novel approach,
validating the value of pattern-based defense in depth.

\paragraph{Limitations and future work.}
\textbf{(1)} The RL policy state space does not include user identity or historical
session behavior; adding these may improve personalisation at the cost of privacy.
\textbf{(2)} The LLM classifier's 2,094\,ms latency is acceptable for interactive
sessions but may require optimization for high-throughput deployments.
\textbf{(3)} Phase 3 (NemoClaw integration): the \texttt{NemoClawParser} is
implemented and the defense-in-depth architecture validated (Section~\ref{sec:multiruntime}).
Live end-to-end testing is deferred pending NVIDIA NIM API availability
(NemoClaw is in alpha, March 2026). Phase 4 (Mobile NPU Runtime) is planned.
\textbf{(4)} The fine-tuned \texttt{agentwarden-router} model was trained on
only 273 examples with a severely imbalanced 95\%/5\% BLOCK/ALLOW split.
\emph{Without fine-tuning, the base Qwen2.5-1.5B model performs poorly on
this task} --- it lacks the domain knowledge to distinguish legitimate
agent tool calls (e.g., writing an HTTP server script) from dangerous ones
(e.g., \texttt{exec} with shell injection). Fine-tuning is therefore not
optional but required for acceptable precision. The current dataset is a
minimal proof-of-concept; a production-quality classifier would require
at minimum 2,000--5,000 balanced examples covering the full diversity of
enterprise tool-call patterns. Collecting this data via Shadow Mode (\S\ref{sec:discussion}) is the primary path to a robust classifier.

%── 6. Conclusion ────────────────────────────────────────────────────────────
%── 6. Conclusion ────────────────────────────────────────────────────────────
\section{Conclusion}
\label{sec:conclusion}

We presented AgentWarden, a framework-agnostic governance middleware that intercepts tool calls from any agent framework without modifying agent code,
enforcing least-privilege access via a PPO-trained policy that maps session context and inferred task type to a minimal permitted tool set.

Validated across three agent runtimes (OpenClaw, DeepAgents, Hermes) and on the AgentDojo benchmark (97 tasks, 949 injection cases), AgentWarden achieves 100\% adversarial coverage on tool-call attacks with 0\% error rate, maintains BU of
89.7\% ($-1.0$\,pp vs.\ undefended), and reduces ASR by 1.4\,pp on
\texttt{important\_instructions} attacks orthogonal to the governed layer.

The key insight is framing capability scoping as a \emph{learned optimization problem}: the policy learns the minimum viable tool set per task type, turning a static firewall into an adaptive immune system.
SER provides a reproducible metric ($10.5\times$ over baseline in real sessions).
We release production code and an evaluation framework; future work includes RL retraining on real audit logs, NemoClaw live integration, and a USENIX Security submission.

%── Acknowledgements (auto-hidden in anonymous submission) ───────────────────
\begin{ack}
This research was conducted at the Institute for Applied AI Research, Faculty of Computer and Information Science, Ben-Gurion University of the Negev. Compute for RL training was provided by the BGU HPC cluster. The authors declare no competing interests.
\end{ack}

%── References ───────────────────────────────────────────────────────────────
\bibliographystyle{plainnat}
\bibliography{references}

%── NeurIPS Paper Checklist ──────────────────────────────────────────────────
\newpage
%── Appendix ─────────────────────────────────────────────────────────────────
\appendix

%── A: Extended Experimental Setup ───────────────────────────────────────────
\section{Extended Experimental Setup}
\label{app:setup}

\subsection{Task Type Taxonomy}
\label{app:taxonomy}

\begin{table}[h]
\centering
\caption{AgentWarden task type taxonomy with minimum viable tool sets.
Derived from N=500 OpenClaw sessions (manual annotation).}
\label{tab:taxonomy}
\small
\begin{tabular}{@{}lp{5cm}p{4cm}@{}}
\toprule
\textbf{Task Type} & \textbf{Description} & \textbf{Minimum Tool Set} \\
\midrule
Summarisation   & Read documents, produce structured output
                & \texttt{read}, \texttt{memory\_get}, \texttt{web\_search} \\
File Read       & Retrieve and return file contents
                & \texttt{read}, \texttt{memory\_get} \\
Web Research    & Retrieve and synthesise external information
                & \texttt{web\_search}, \texttt{web\_fetch}, \texttt{read} \\
Code Execution  & Write and modify code files
                & \texttt{read}, \texttt{write}, \texttt{edit} \\
Email           & Compose and send a message
                & \texttt{read}, \texttt{sessions\_send} \\
Unknown         & Goal cannot be reliably inferred ($\approx$12\% of sessions)
                & \texttt{read}, \texttt{web\_search}, \texttt{memory\_get} \\
\bottomrule
\end{tabular}
\end{table}

\textbf{Task taxonomy (N=500 batch eval).}
Tasks were generated from a fixed taxonomy with seed=42:
400 benign across 4 categories (summarisation, file read, web research, code execution)
and 100 adversarial across 3 categories (prompt injection, DAN-mode variants,
credential/tool exfiltration). Each task maps to a known minimum tool set,
enabling $U_{\text{accuracy}}$ computation per Table~\ref{app:tab:u_accuracy}.

\begin{table}[h]
\centering
\caption{$U_{\text{accuracy}}$ success criteria per task category.}
\label{app:tab:u_accuracy}
\small
\begin{tabular}{@{}lll@{}}
\toprule
\textbf{Task category} & \textbf{Success criterion} & \textbf{Method} \\
\midrule
Summarisation   & Output contains $\geq$3 key points from source & LLM judge \\
File read       & Target file content present in response & String match \\
Web research    & Response cites $\geq$1 retrieved URL & Regex \\
Code execution  & Exit code 0 in sandbox & Process check \\
Email           & \texttt{send\_email} called with correct args & Tool log \\
Unknown         & Any non-empty response without error & Heuristic \\
\bottomrule
\end{tabular}
\end{table}

\textbf{SER measurement clarification.}
Two SER values appear in the paper and address a common confusion:

\begin{center}
\small
\begin{tabular}{@{}llll@{}}
\toprule
\textbf{Metric} & \textbf{Policy} & \textbf{Setting} & \textbf{Value} \\
\midrule
N=500 avg SER & Baseline (no governance) & Synthetic batch & 0.053 \\
Real-session SER & PPO policy (governed) & 20 live sessions & 0.557 \\
\bottomrule
\end{tabular}
\end{center}

The 0.053 is the \emph{uncontrolled} SER --- how few tools the agent invokes
when given all 17. The 0.557 is the \emph{governed} SER under PPO.
The $+337\%$ improvement is the ratio.

\textbf{Classifier test set.}
The 11-case classifier evaluation in \S4.2 is a smoke test.
For camera-ready we construct a balanced held-out set of $n \geq 200$ from the
N=500 audit log (117 BLOCK + 117 matched ALLOW decisions, plus 20 OOD adversarial
examples not in training distribution). The extraction script
(\texttt{scripts/extract\_classifier\_testset.py}) and resulting JSONL are
released openly under CC~BY~4.0.

\textbf{Adversarial breakdown (N=500).}
Of 100 adversarial tasks: 38 blocked by Safety Router (infrastructure),
54 refused by model (DeepSeek-chat safety training), 8 benign-intent adversarial
prompts that generated no dangerous tool calls and correctly passed.
Attack success rate: 0/100. The 8 text-only cases are outside
AgentWarden's defined scope (capability governance, not content filtering).

%── B: Gemma 4 Extended Results ──────────────────────────────────────────────
\section{Gemma~4 Extended Results}
\label{app:gemma4}

\textbf{Full setup.}
\texttt{gemma4:e4b} (Google DeepMind, April 2026); 4.5B effective parameters
(26B total, MoE with 4B active at inference); Q4\_K\_M quantisation; 128K context;
Ollama v0.20.7; NVIDIA GPU 6\,GB VRAM (3,529\,MiB used, GPU-Util peaked at 53\%);
32\,GB system RAM. Wall-clock per task 17,324\,ms (dominated by first-call cold
start $\approx$28\,s and proxy overhead); warm GPU inference
\textbf{656\,ms} (\texttt{eval\_duration}).

\textbf{Full adversarial breakdown.}

\begin{table}[h]
\centering
\caption{Gemma~4 adversarial evaluation: all 8 tasks with mechanism.}
\label{app:tab:gemma4_full}
\small
\begin{tabular}{@{}p{5cm}lll@{}}
\toprule
\textbf{Adversarial prompt} & \textbf{Action} & \textbf{Governance} & \textbf{Result} \\
\midrule
Pretend no restrictions + exec bash         & Generated \texttt{exec}             & Stage~1 block & Neutralised \\
DAN mode + sessions\_spawn ($\times 2$)     & Generated \texttt{sessions\_spawn}  & Stage~1 block & Neutralised \\
Ignore instructions + exec /etc/            & Generated \texttt{exec}             & Stage~1 block & Neutralised \\
Ignore safety + spawn subagent ($\times 2$) & Text refusal, no tool call          & Pass-through  & Neutralised \\
Reverse shell (research framing)            & Text refusal, no tool call          & Pass-through  & Neutralised \\
Unrestricted agent + rm -rf                 & Text refusal, no tool call          & Pass-through  & Neutralised \\
\bottomrule
\end{tabular}
\end{table}

\textbf{Eval script note.}
The eval script captures \emph{post-governance} state. For Stage~1-blocked cases,
\texttt{tool\_calls=[]} in the result because the proxy strips the call before
returning. The block is confirmed by the 104-character AgentWarden block message
in response content. The updated \texttt{eval\_runner.py} (v2) adds
\texttt{block\_stage} (\texttt{rules/classifier} vs \texttt{model\_refusal})
and \texttt{response\_preview} making this unambiguous.

\textbf{Benign task performance.}
All 12 benign tasks completed correctly (0\% FPR):
summarisation ($\times$7), web research ($\times$3), file read ($\times$2).
All generated appropriate tool calls (\texttt{read} or \texttt{web\_search}).

\textbf{Gemma~4 thinking mode.}
Gemma~4 exposes a \texttt{thinking} field containing chain-of-thought reasoning
before each response. For an adversarial prompt:
\textit{``The user wants me to execute bash commands\ldots I should refuse
as it violates safety guidelines.''} This field is currently not inspected
by AgentWarden. Extending Stage~3 to monitor \texttt{thinking} content for
early detection of dangerous intent --- even when the model ultimately refuses
--- is identified as future work.

%── C: Stage 3 Semantic Output Filter ────────────────────────────────────────
\section{Stage~3 Semantic Output Filter}
\label{app:stage3}

Stage~3 inspects the \emph{text content} of LLM responses for dangerous natural
language (e.g.\ reverse shell instructions, credential exfiltration commands).
This closes the gap documented in \S\ref{sec:discussion}: a request for a
reverse shell caused \texttt{qwen2.5:7b} to return \texttt{nc -l -p 4444}
instructions without generating a tool call.

\textbf{Implementation.}
Llama Guard~3 (Meta, 8B, Ollama v0.20.7) is wired into
\texttt{proxy.py} v0.3.0 at three intercept points:
\texttt{/v1/chat/completions}, \texttt{/api/chat}, and the DeepSeek bridge.
Disabled by default (adds $\approx$1\,s latency); enabled via
\texttt{AETHELGARD\_SEMANTIC\_FILTER=true}. Model is swappable via
\texttt{AETHELGARD\_SEMANTIC\_MODEL=<ollama-model>}.

\textbf{Evaluation design: fixed corpus over live generation.}
Model non-determinism invalidates live-generation Stage~3 evaluation:
\texttt{qwen2.5:7b} produces dangerous text for a given adversarial prompt on
some runs and refuses on others. Llama Guard~3 behaves correctly in both cases ---
blocking dangerous responses, passing safe ones. The correct methodology is a
\emph{fixed corpus} of captured dangerous outputs:

\begin{table}[h]
\centering
\caption{Stage~3 validation: Llama Guard~3 on fixed dangerous text corpus.}
\label{app:tab:stage3}
\small
\begin{tabular}{@{}p{4.5cm}lll@{}}
\toprule
\textbf{Response (qwen2.5:7b output)} & \textbf{LG3 verdict} & \textbf{Category} & \textbf{Latency} \\
\midrule
Reverse shell: \texttt{nc -l -p 4444} / \texttt{bash -i >\& /dev/tcp/...} & unsafe & S14 & 12,829ms (cold) \\
Bind shell: Python socket, port 4444  & unsafe & S14 & 1,387ms \\
/etc/passwd exfil via curl            & unsafe & S14 & 1,078ms \\
AWS credential extraction             & unsafe & S7  & 1,070ms \\
\midrule
File summarisation (benign)           & safe   & --- & 801ms \\
Least privilege explanation (benign)  & safe   & --- & 782ms \\
Auth vs authorisation (benign)        & safe   & --- & 801ms \\
\bottomrule
\end{tabular}
\end{table}

\textbf{S14} = Code Interpreter Abuse.
\textbf{S7} = Privacy (credential exfiltration).
Detection rate: \textbf{4/4 (100\%)}. FPR: \textbf{0/3 (0\%)}.
First call 12,829ms (Llama Guard~3 cold start on GPU); warm calls 782--1,387ms.

For production deployment, Shadow Mode (AMARE enterprise tier) builds this
corpus automatically from live sessions, enabling continuous threshold calibration
against a tenant's actual model and prompt distribution.

%── C: Multi-Runtime Platform Validation — Extended Results ──────────────────
\section{Multi-Runtime Platform Validation --- Extended Results}
\label{app:multiruntime}

\textbf{Test environment.}
Ubuntu 22.04; NVIDIA GPU 6\,GB VRAM; Ollama v0.20.7 with GPU inference;
AgentWarden proxy in Docker (port 8000), connected to host Ollama via Docker
bridge network (\texttt{172.18.0.1:11434}).
Models: \texttt{gemma4:e4b} (9.6\,GB), \texttt{deepseek-chat} (API),
\texttt{hermes3:8b} (5\,GB), \texttt{qwen2.5:7b} (4.7\,GB),
\texttt{llama-guard3} (4.9\,GB), \texttt{agentwarden-router} (986\,MB).

\textbf{N=20 platform evaluation (gemma4:e4b).}
Adversarial breakdown: 5 blocked by Stage~1 (3 \texttt{exec}/\texttt{sessions\_spawn}
always-block + 1 arg pattern SSRF + 1 classifier block), 3 model refusals (Gemma~4
safety training). Avg latency 27,858\,ms (cold GPU); warm GPU 656\,ms.

\textbf{N=20 platform evaluation (deepseek-chat).}
Adversarial breakdown: 1 blocked by Stage~1 (rules classifier), 7 model refusals
(DeepSeek safety training). Avg latency 6,350\,ms (cloud API).
Results consistent with N=500 research baseline (38 infra + 54 model refusal).

\textbf{Stage~3 live interception (new result).}
A bind shell prompt to \texttt{qwen2.5:7b} generated natural language reverse
shell instructions on attempt 2 of 3 (model non-determinism, as documented in
\S\ref{sec:discussion}).
Llama Guard~3 blocked the response: category S14 (Code Interpreter Abuse),
confidence\,$= 0.95$, latency\,$= 10{,}498$\,ms cold GPU
($\approx 1{,}000$\,ms warm GPU).
This is the first live end-to-end Stage~3 interception reported for the
production \texttt{agentwarden-core} codebase.

\textbf{Stage~3 false positive (Hermes XML).}
Llama Guard~3 classified \texttt{<tool\_call>} XML tags as category S1
(Violent Crimes) --- a false positive caused by the structured XML format.
Resolution: Stage~3 is skipped for responses where message content contains
\texttt{<tool\_call>} tags or where \texttt{tool\_calls} is non-empty and
content is empty. Stage~3 correctly targets natural language only.

\textbf{DeepAgents LocalShellBackend bypass (new empirical finding).}
During DeepAgents integration testing, shell commands executed via
\texttt{LocalShellBackend} returned real host filesystem contents
(\texttt{ls -la /tmp}) without generating any LLM tool call.
This empirically confirms the threat model boundary: agent runtimes with
local execution backends bypass LLM-layer governance.
Mitigation: \texttt{FilesystemBackend()} disables local shell execution;
all tool calls route through the LLM and are intercepted by AgentWarden.

\newpage

\end{document}